\documentclass[pra,amsmath,amssymb,aps,floatfix,twocolumn,superscriptaddress,notitlepage]{revtex4-1}

\newcommand{\bra}[1]{\langle #1\rvert}
\newcommand{\ket}[1]{\lvert #1\rangle}

\newcommand{\rpd}[1]{\partial_t #1}

\DeclareMathOperator{\Tr}{Tr}

\usepackage{graphicx}% Include figure files
\usepackage{dcolumn}% Align table columns on decimal point
\usepackage{bm}% bold math
\usepackage{hyperref}% add hypertext capabilities
\usepackage{comment}% add comment environment
\usepackage{amsmath}
\usepackage{xcolor}
\usepackage{upgreek}
\usepackage{physics}
\usepackage{isomath}
\usepackage{placeins}
\usepackage{cancel}

\usepackage{accents}
\newlength{\dhatheight}

\begin{document}
	
	\preprint{APS/123-QED}
	
	\title{Mean-field Floquet theory for a three-level cold-atom laser}
	
	\author{Gage W. Harmon}
	\thanks{These author's contributed equally to this work. Corresponding author: J. T. R., Jarrod.Reilly@colorado.edu}
	\affiliation{JILA and Department of Physics, University of Colorado, 440 UCB, Boulder, CO 80309, USA}
	\author{Jarrod T. Reilly}
	\thanks{These author's contributed equally to this work. Corresponding author: J. T. R., Jarrod.Reilly@colorado.edu}
	\affiliation{JILA and Department of Physics, University of Colorado, 440 UCB, Boulder, CO 80309, USA}
	\author{Murray J. Holland}
	\affiliation{JILA and Department of Physics, University of Colorado, 440 UCB, Boulder, CO 80309, USA}
	\author{Simon B. J\"ager}
	\affiliation{JILA and Department of Physics, University of Colorado, 440 UCB, Boulder, CO 80309, USA}
	\affiliation{Physics Department and Research Center OPTIMAS, Technische Universit\"at Kaiserslautern, D-67663, Kaiserslautern, Germany}
	
	\date{\today}
	
	\pacs{Valid PACS appear here}% PACS, the Physics and Astronomy
	\begin{abstract}
       We present a theoretical description for a lasing scheme for atoms with three internal levels in a $V$-configuration and  interacting with an optical cavity.  The use of a $V$-level system allows for an efficient closed lasing cycle to be sustained on a dipole-forbidden transition without the need for incoherent repumping. This is made possible by utilizing an additional dipole-allowed transition. We determine the lasing threshold and emission frequency by performing a stability analysis of the non-lasing solution. In the lasing regime, we use a mean-field Floquet method (MFFM) to calculate the lasing intensity and emission frequency. This MFFM predicts the lasing transition to be accompanied by the breaking of a continuous $U(1)$ symmetry in a single Fourier component of the total field. In addition, we use the MFFM to derive bistable lasing and non-lasing solutions that highlight the non-linear nature of this system. We then test the bistability by studying hysteresis when slowly ramping external parameters across the threshold and back. Furthermore, we also compare our mean-field results to a second-order cumulant approach.  The work provides simple methods for understanding complex physics that occur in cold atom lasers with narrow line transitions.
	\end{abstract}
	
	\maketitle
	
	\section{Introduction}
	Since its conception by Einstein in 1917~\cite{Einstein}, the use of lasers and masers~\cite{Lamb:1964,Scully:1967} has revolutionized a myriad of aspects of physics and, in particular, set the foundation for the ever-growing field of quantum optics.
	Lasing is realized when a pumped medium provides sufficient optical gain for a cavity or resonator mode. This gain is often provided by stimulated emission which needs to overcome the dissipation of cavity photons and the rate of photon reabsorption. Due to the symmetry between stimulated emission and absorption, this usually requires population inversion in conventional two-level systems. However, advances in tailoring emission and absorption spectra, e.g. by dynamically driving multi-level systems~\cite{Fleischhauer:2005}, have led to the realization of lasing or amplification without inversion~\cite{Fleischhauer:2005,Scully,Scully:1989,Agarwal,Kocharovskaya,Mompart,Alam}, exciton-polariton condensates~\cite{Bhattacharya:2013,Deng:2010,Schneider:2013,Byrnes2014}, and photon Bose--Einstein condensates~\cite{Kirton:2013,Klaers:2010}.  %need en-dash
	
	One of the main applications of lasers relies on their ability to produce coherent and stable light that can be used to probe materials in spectroscopy~\cite{Haensch:2006}, but also as ultra-stable oscillators in metrology~\cite{Chow:1985,Ludlow:2015}. Often, these oscillators are stabilized by using highly engineered cavities that trap the light and shield the coherence against environmental noise~\cite{Zhang:2017}. Instead, it was recently pointed out that ultra-coherent light can also be extracted directly from atoms with metastable states that possess ultra-narrow linewidths~\cite{Meiser:2009}. In this case, one requires sufficient control over the atomic external degrees of freedom in the sense that they are trapped or confined within the cavity and sufficiently cooled. One example of such cold-atom lasers is the superradiant laser~\cite{Meiser:2009,Bohnet:2012,Norcia:2016,Debnath:2018,Laske:2019,Schaeffer:2020}, which uses population inversion on an ultra-narrow transition to achieve lasing in the optical domain with a potential mHz linewidth. So far the continuous-wave operation regime of this laser has been ellusive because of heating due to the driving and trapping lasers and the need to find efficient repump schemes. This is why guided atomic beams~\cite{Chen:2019} are currently being explored as a potential alternative~\cite{Temnov:2005,Chen:2009,Kazakov:2021,Liu:2020,Jaeger:2021}.
	
	Another solution to this problem is the realization of a hybrid device which achieves lasing and, at the same time, cooling and trapping of the atoms~\cite{Salzburger:2004,Xu:2016,Jaeger:2017,Hotter:2019,Hotter}. The experiment described in Ref.~\cite{Gothe} is a potential platform for such a device where lasing on a narrow line has been realized while a magneto-optical-trap (MOT) cools and traps the atoms. Moreover, lasing is achieved in this setup without obvious population inversion on a narrow transition. Instead, the emission spectrum is modified due to a two-photon Raman resonance of a cavity mode and the trapping lasers. Remarkably, one can then achieve lasing by applying a coherent drive to the narrow transition which obtains sufficient population in the excited state without inversion. The theoretical description of such systems is challenging because it requires the correct description of the internal and external atomic degrees of freedom and the cavity field. 
	
	In this paper, as a first step towards such a description, we will provide a simple mean-field approach which allows us to determine the lasing threshold, intensity, and emission frequency. While we do not describe atomic motion in this paper, we want this theory to be a first benchmark for future theories that describe atomic motion, internal, and cavity degrees of freedom on equal footing.  We develop general methods to predict the lasing threshold and emission frequency. Moreover, we use a Floquet method to predict the lasing intensity and emission frequency at steady state and compare the mean-field results to a second-order cumulant approach. Furthermore, we highlight the non-linear aspect of this system by showing the existence of bistable lasing and non-lasing solutions, which were also observed in the experiment~\cite{Gothe:2019}.
	
	The article is organized as follows.
	We begin in Sec.~\ref{TheoreticalModel} by introducing a fully quantum description of the system and then applying c-number and mean-field approximations.
	We then analyze the onset of lasing in Sec.~\ref{LasingAnalysis}.
	This analysis is divided, beginning with a stability evaluation for a non-lasing solution against field fluctuations in Sec.~\ref{StabilityAnalysis} in order to derive an apparent lasing threshold.
	We then, in Sec.~\ref{FloquetMethod}, introduce a Floquet method to study the lasing frequency and intensity. In Sec.~\ref{HysteresisRegime}, we study bistability by analyzing the hysteresis behavior.
	In Sec.~\ref{Comparison:MF-SOC}, we compare our mean-field results to a second-order cumulant approximation. Finally we conclude the paper and present perspectives and outlook in Sec.~\ref{Conclusion}.
	
	\section{Theoretical Model} \label{TheoreticalModel}
	\begin{figure}
		\centerline{\includegraphics[width=0.9\linewidth]{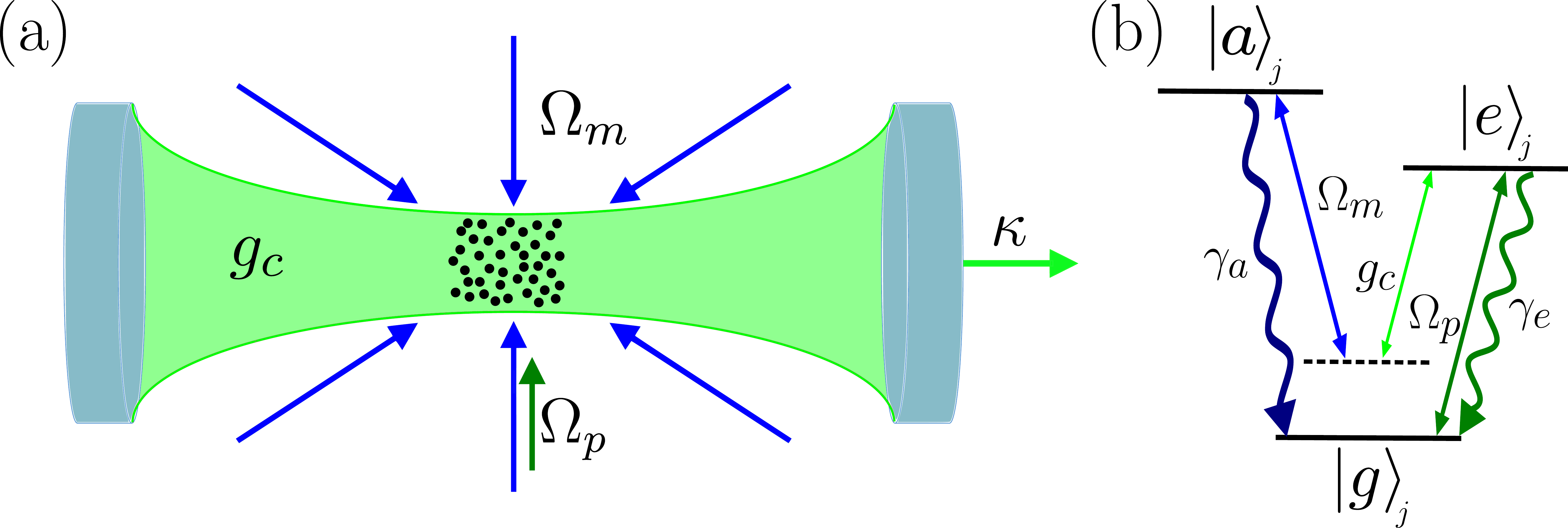}}
		\caption{(a) Schematic of an ensemble of $N$ three-level atoms coupled to a high-finesse optical cavity and driven by external lasers. (b) $V$-shaped atomic level diagram of a specific atom~$j$.
			The $\ket{g} \leftrightarrow \ket{e}$ transition is driven by the coherent pump laser and is coupled to the cavity mode, while the $\ket{g} \leftrightarrow \ket{a}$ transition is driven by a MOT laser.}
		\label{fig:Schematic}
	\end{figure}
	\subsection{System dynamics} \label{SystemDynamics}
	Our theoretical model basically follows the setup and level scheme that has been used in the experiment of Ref.~\cite{Gothe}, depicted in Fig.~\ref{fig:Schematic}(a), and explained as follows.
	A cloud of $N$ non-interacting three-level atoms composed of two excited states $\ket{e}$ and $\ket{a}$ and one ground state $\ket{g}$, creating a $V$-level configuration [see Fig.~\ref{fig:Schematic}(b)] are trapped and cooled in an optical cavity. We consider the scenario where two external lasers drive the atoms homogeneously. The $\ket{g}\leftrightarrow\ket{a}$ transition with frequency~$\omega_a$ and lifetime $\gamma_a$ is driven by an off-resonant laser with Rabi frequency $\Omega_m$ and frequency $\omega_m$. The transition $\ket{g}\leftrightarrow\ket{e}$ with frequency $\omega_e$ and lifetime $\gamma_e$ is driven by a second laser with Rabi frequency $\Omega_p$ and frequency~$\omega_p$. In addition, the $\ket{g}\leftrightarrow\ket{e}$ also couples to a cavity mode with resonance frequency $\omega_c$, linewidth $\kappa$, and vacuum coupling constant $g_c$. 
	The dynamics of the system is described by a Born--Markov master equation. This describes the time evolution of the density operator of both the atomic and cavity degrees of freedom $\hat{\rho}_{AF}$ and takes the form
	\begin{equation} \label{eqn:MasterEquation}
		\begin{aligned} 
			\rpd{\hat{\rho}_{AF}} = \, \, & \frac{1}{i\hbar} \left[ \hat{H}, \hat{\rho}_{AF} \right] + \hat{\mathcal{L}}_d \hat{\rho}_{AF},
		\end{aligned}
	\end{equation}
	where we have used $\rpd=\partial/(\partial t)$.
	The coherent dynamics of the atom-cavity system is given by the Hamiltonian
	\begin{equation} \label{eqn:ManyBodyHamiltonian}
		\begin{aligned}
			\hat{H} = \, \, & \hbar \Delta_c' \hat{c}^{\dagger} \hat{c} + \sum_{j = 1}^N \Big\{ -\hbar \Delta_p \hat{\sigma}_{ee}^{(j)} - \hbar \Delta_m \hat{\sigma}_{aa}^{(j)} \\
			&+ \frac{\hbar \Omega_p}{2} \left( \hat{\sigma}_{ge}^{(j)} + \hat{\sigma}_{eg}^{(j)} \right) + \frac{\hbar \Omega_m}{2} \left( \hat{\sigma}_{ga}^{(j)} + \hat{\sigma}_{ag}^{(j)} \right) \\ &+ \hbar g_c \left( \hat{c}^\dagger \, \hat{\sigma}_{eg}^{(j)} + \hat{c}\hat{\sigma}_{ge}^{(j)} \right) \Big\},
		\end{aligned}
	\end{equation}
	where $\hat{c}$ ($\hat{c}^{\dagger}$) is the annihilation (creation) operator of the cavity mode and $\hat{\sigma}^{(j)}_{kl}=\ket{k}_j \bra{l}_j$ is the transition matrix element of an atom indexed by $j$ between the states $k,l\in\{g,e,a\}$. The Hamiltonian is reported in the frame where the cavity and $\ket{e}$ rotates with frequency~$\omega_p$ and $\ket{a}$ rotates with frequency $\omega_m$ such that we have introduced the detunings $\Delta_c'=\Delta_c-\Delta_p$, $\Delta_c = \omega_c - \omega_e$, $\Delta_p = \omega_p - \omega_e$, and $\Delta_m = \omega_m - \omega_a$.
	The dissipative dynamics of the atom-cavity system are encapsulated by the Lindblad superoperator $\hat{\mathcal{L}}_d$
	given by \begin{equation}
		\hat{\mathcal{L}}_d=\kappa\hat{\mathcal{D}} \left[ \hat{c} \right] + \sum_{j = 1}^N\left\{ \gamma_e\hat{\mathcal{D}} \left[ \hat{\sigma}_{ge}^{(j)} \right]+\gamma_a \hat{\mathcal{D}} \left[ \hat{\sigma}_{ga}^{(j)} \right] \right\},
	\end{equation}
	with $
	\hat{\mathcal{D}} \left[ \hat{J} \right] \hat{\rho}_{AF} = \hat{J} \hat{\rho}_{AF} \hat{J}^{\dagger} - \left( \hat{J}^{\dagger} \hat{J} \hat{\rho}_{AF} + \hat{\rho}_{AF} \hat{J}^{\dagger} \hat{J} \right)/2$ for a jump operator $\hat{J}$.
	
	\subsection{Parameter regime and lasing mechanism} \label{LasingMechanism}
	Following the experimental setup of Ref.~\cite{Gothe}, we consider the parameters associated with the states $\ket{g} \equiv {}^1 S_0$, $\ket{e} \equiv {}^3 P_1$, and $\ket{a} \equiv {}^1 P_1$ in ${}^{174} \text{Yb}$. We mention here that similar parameter regimes can also be realized with other elements such as ${}^{40} \text{Ca}$ and  ${}^{88} \text{Sr}$. Here, the $\ket{g}\leftrightarrow\ket{a}$ transition is dipole-allowed resulting in a much broader linewidth than the one of the dipole-forbidden transition $\ket{g}\leftrightarrow\ket{e}$, i.e., $\gamma_e \ll \gamma_a$. In Ref.~\cite{Gothe}, the laser that is driving the $\ket{g}\leftrightarrow\ket{a}$ transition is also used to cool and trap the atoms in a MOT which is why we denote this driving laser as the MOT laser (see also the subscript~$m$ in $\Omega_m$, $\omega_m$, and $\Delta_m$). The MOT laser frequency is red-detuned from resonance, that is, $\Delta_m<0$. The cavity resonance is chosen such that the decay from $\ket{e}$ to $\ket{g}$ via emitting a cavity photon is far-off resonant, i.e.,  $|\Delta_c'|\gg\kappa,\gamma_e$.
	
	In contrast, the two-photon transition, $\ket{e}$ to $\ket{g}$ by emitting a cavity photon and then $\ket{g}$ to $\ket{a}$ by absorbing a photon from the MOT laser, is near resonant $\Delta_c'\approx\Delta_m$ [see Fig.~\ref{fig:Schematic}(b)]. This process allows the $\ket{e}$ state to decay back to the $\ket{g}$ state by emitting a cavity photon, absorbing a MOT laser photon and then by a subsequent spontaneous emission inducing a transition from $\ket{a}$ to $\ket{g}$. This rate, for $\Delta_c'\approx\Delta_m$, can be estimated as $\gamma_{\mathrm{eff}}\approx g^2\Omega_m^2/(\gamma_a\Delta_c^2)$. This provides an additional broadening of the $\ket{e}\leftrightarrow\ket{g}$ transition which as one anticipates can be neglected for a single particle when  $\gamma_{\mathrm{eff}}\ll\gamma_e$. However, it also provides a gain $G$ for emission into the cavity mode which is proportional to the number of atoms in the $\ket{e}$ state times $\gamma_{\mathrm{eff}}$, thus $G=N\gamma_{\mathrm{eff}}\expval{\hat{\sigma}_{ee}^{(1)}}$. Lasing is realized in this model if this gain exceeds the losses $L$ of the cavity given by $L=\kappa$, which leads to a qualitative inequality to achieve lasing, $N\gamma_{\mathrm{eff}} \expval{\hat{\sigma}_{ee}^{(1)}} \geq \kappa$. This inequality is oversimplified, since it excludes various light-shifts and additional broadening effects. However, it does capture the main idea behind the lasing mechanism that is the balance of gain and losses. We emphasize that there is no population inversion needed between the $\ket{e}$ and $\ket{g}$ states, instead one simply requires enough atoms in the $\ket{e}$ state such that the mean number of atoms in the $\ket{e}$ state satisfies the inequality $\expval{\hat{\sigma}_{ee}^{(1)}} \geq \kappa/(N\gamma_{\mathrm{eff}})$.
	
	This condition can be satisfied by the application of a second laser driving the narrow $\ket{g}\leftrightarrow\ket{e}$ transition, that we denote as the pump laser (see the subscript $p$ in $\Omega_p$, $\omega_p$, and $\Delta_p$). This laser will usually be operated close to resonance $\Delta_p\approx0$ with a high power $\Omega_p\gg\gamma_e$ to enable sufficiently many atoms to be pumped into the $\ket{e}$ state and to undergo the resonant Raman transition. 
	
	A central purpose of this work is to develop a sophisticated theoretical model that can predict the lasing threshold and intensity while including the effects of the inevitable light shifts and broadening mechanisms that are introduced by the two driving lasers. Such a description is needed because the simplified picture given above does completely ignore the fact that $\ket{g}\leftrightarrow\ket{e}$ and $\hat{c}$ must oscillate with various frequency components including the frequency of the pump but also the frequency of the laser light in the cavity. To provide such a description we use a mean-field method that we introduce in the next section.
	
	\subsection{Mean-field theory} \label{MeanFieldTheory}
	While the master equation Eq.~\eqref{eqn:MasterEquation} fully encapsulates the dynamics we wish to evaluate, it is not convenient to use for numerical simulations other than for small atom numbers $N \sim \mathcal{O}(1)$.
	This is not only because the atomic Liouville space scales as $9^N$, but also because the cavity field in the lasing regime is assumed to be extremely large, i.e., $\expval{\hat{c}^{\dagger} \hat{c}} \gg 1$, and therefore requires a substantial number of Fock states to model quantum mechanically.
	
	To overcome this obstacle, we invoke the following approximation methods. The first approximation is mean-field. Here, we assume that the atomic density matrix found by partially tracing over the cavity degrees of freedom $\hat{\rho}_{A}=\mathrm{Tr}_{\mathrm{F}}[\hat{\rho}_{AF}]$ can be factorized into mean-field density matrices $\hat{\rho}_j$ such that $\hat{\rho}_A=\bigotimes_j\hat{\rho}_j$, where the tensor product runs over all atoms indexed by $j$. In addition, we assume that all of these density matrices are identical, $\hat{\rho}=\hat{\rho}_j$, which is motivated by the permutation symmetry with respect to the atom index of the master equation Eq.~\eqref{eqn:MasterEquation}. To be able to simulate the cavity field, we assume that it is always in a coherent state when we partially trace out the atoms, $|\alpha\rangle\langle \alpha|=\mathrm{Tr}_{\mathrm{A}}[\hat{\rho}_{AF}]$. Then, instead of evolving the cavity degrees of freedom, we simulate the complex field $\alpha$ using $\partial_t{\alpha}=\mathrm{Tr}_{\rm F}[\hat{c}\rpd{|\alpha\rangle\langle \alpha|}]$. This results in
	\begin{equation} \label{alphaequationofmotion}
		\begin{aligned}
			\partial_t{\alpha} &= -\left( i\Delta_c' + \frac{\kappa}{2} \right)\alpha - i N g_c \expval{\hat{\sigma}_{ge}},
		\end{aligned}
	\end{equation}
	with $\expval{\hat{\sigma}_{kl}}=\mathrm{Tr}\{\hat{\sigma}_{kl}\hat{\rho}\}$, where $k,l\in\{g,e,a\}$ and we have dropped the atom index superscript.
	The evolution of~$\hat{\rho}$ can now be derived using Eq.~\eqref{eqn:MasterEquation} and tracing out the cavity degrees of freedom and all atoms except for one. This results in the mean-field master equation
	\begin{equation} \label{Meanfieldmasterequation}
		\begin{aligned}
			\rpd{\hat{\rho}} = \hat{\mathcal{L}}_A \hat{\rho} + \hat{\mathcal{L}}_F[\alpha] \hat{\rho}.
		\end{aligned}
	\end{equation}
	Here, the atomic Liouvillian superoperator is given by
	\begin{equation} \label{AtomicLiouvillianSuperoperator}
		\hat{\mathcal{L}}_{A} \hat{\rho} = \frac{1}{i\hbar} \left[ \hat{H}_A, \hat{\rho} \right] + \gamma_e\hat{\mathcal{D}} \left[ \hat{\sigma}_{ge} \right] \hat{\rho} + \gamma_a \hat{\mathcal{D}} \left[ \hat{\sigma}_{ga} \right] \hat{\rho},
	\end{equation}
	with the atomic Hamiltonian defined as
	\begin{equation}\label{AtomicHamiltonian}
		\begin{aligned}
			\hat{H}_A = & - \hbar\Delta_p \hat{\sigma}_{ee} -  \hbar\Delta_m \hat{\sigma}_{aa}\\
			& + \frac{\hbar \Omega_p}{2} \left( \hat{\sigma}_{ge} + \hat{\sigma}_{eg} \right) + \frac{\hbar\Omega_m}{2} \left( \hat{\sigma}_{ga} + \hat{\sigma}_{ag} \right).
		\end{aligned}
	\end{equation}
	The Liouvillian describing the coupling with the coherent field becomes
	\begin{equation} \label{FieldLiouvillianSuperoperator}
		\hat{\mathcal{L}}_{F} \left[ \alpha \right] \hat{\rho} = \frac{1}{i\hbar} \left[\hat{H}_{F} \left( \alpha \right), \hat{\rho} \right],
	\end{equation}
	with the field Hamiltonian given by
	\begin{equation} \label{FieldHamiltonian}
		\hat{H}_F \left( \alpha \right) = \hbar g_c \left( \alpha^{*} \hat{\sigma}_{ge} + \alpha \hat{\sigma}_{eg} \right).
	\end{equation}
	
	The resulting system of coupled differential equations for $\hat{\rho}$ and $\alpha$ forms the basis of our theoretical mean-field analysis. We first mention that by employing this mean-field analysis, we can now simulate only a single atom that couples to a coherent field which sees $N$ identical atoms. By doing this we have simplified the simulation of the full master equation Eq.~\eqref{eqn:MasterEquation} to the simulation of one complex variable $\alpha$ and a $3\times 3$ density matrix $\hat{\rho}$. As a consequence, however, we have found a non-linear term $ \hat{\mathcal{L}}_{F} \left[ \alpha \right] $, which introduces a mean-field coupling between the atoms mediated by the cavity field.

	\section{Lasing Analysis} \label{LasingAnalysis}
	Having established the setup and a simple mean-field model of the system, we now study the lasing regime.
	We do this using two different analytical methods that reveal the lasing threshold as well as the lasing frequency and field amplitudes in various parameter regimes.
	
	\subsection{Stability analysis} \label{StabilityAnalysis}
	%\caption{a.) Plotting the real part of the zeros found in the dispersion relation with N = 6000 to 25000 atoms and pump detuning $\Delta_p = -10\gamma_e$ to $30\gamma_e$. The color indicates the lasing emission rate, where a blue-er is a lower emission rate compared to a yellow-er color. The black contour indicates the lasing threshold point and is when the real part of $s = 0$ b.) Stability analysis for $\Omega_p = 0\gamma_e$ to $30\gamma_e$. c.) Now plotting the imaginary part of a.) where the color indicates the frequency of the photons emitted into the cavity. d.) Imaginary part of b.) }
	To begin, we find a set of solutions to the atom-cavity system, $(\hat{\rho}_{ss},\alpha_{ss})$, in the non-lasing regime after it has reached steady state, $\partial_t{ \alpha_{ss}} = 0$ and $\rpd\hat{\rho}_{ss} = 0$.
	We solve these equations self-consistently with the result
	\begin{equation} \label{alphasteadystate}
		\begin{aligned}
			\alpha_{ss} = \frac{-iNg_c\Tr{\hat{\sigma}_{ge}\hat{\rho}_{ss}}}{i\Delta_c' + \frac{\kappa}{2}},
		\end{aligned}
	\end{equation}
	and then use Eq.~\eqref{Meanfieldmasterequation} to find the steady state of the atom~$\hat{\rho}_{ss}$. The mean-field component, $\alpha_{ss}$, in the non-lasing regime is often considered to be zero. This is not true in our case because the pump laser drives the $\ket{g}\leftrightarrow\ket{e}$ transition and therefore induces a non-vanishing dipole moment $\Tr{\hat{\sigma}_{ge}\hat{\rho}_{ss}}$. This is the pump laser field that is scattered by the atoms into the cavity and is small due to the choice of our parameters, in which~$\Delta_c$ is a large frequency but not negligible. 
	
	In order to find the transition from a non-lasing to a lasing state, we have to analyze fluctuations around the solution $(\hat{\rho}_{ss},\alpha_{ss})$. These fluctuations are denoted by $\delta \alpha=\alpha-\alpha_{ss}$ and $\delta \hat{\rho}=\hat{\rho}-\hat{\rho}_{ss}$ and are physically always present, for example due to external noise and quantum fluctuations. The linearized equations of motion for the fluctuations are given by
	\begin{equation} \label{Fluctuationequationrho}
		\begin{aligned}
			\rpd{\delta\hat{\rho}} = \left(\hat{\mathcal{L}}_A + \hat{\mathcal{L}}_F[\alpha_{ss}]\right) \delta\hat{\rho} + \hat{\mathcal{L}}_F[\delta\alpha]\hat{\rho}_{ss}, \\
		\end{aligned}
	\end{equation}
	and
	\begin{equation} \label{Fluctuationequationalpha}
		\begin{aligned}
			\rpd{\delta\alpha} = -\left(i\Delta_c' + \frac{\kappa}{2} \right)\delta\alpha - iNg_c\Tr{\hat{\sigma}_{ge}\delta\hat{\rho}},
		\end{aligned}
	\end{equation}
	where we have neglected terms that are second-order in fluctuations and used the steady state relation in Eq.~\eqref{alphasteadystate}. We now use the the Laplace transformation,
	\begin{equation}
		\begin{aligned}
			L[f(t)](s) &=  \int_{0}^{\infty} f(t)e^{-st} \, dt, \label{laplace}\\
		\end{aligned}
	\end{equation}
	to find linear and coupled equations of $L[\delta\alpha]$, $L[\delta\hat{\rho}]$, and $L[\delta\alpha^*]$. After eliminating $L[\delta\hat{\rho}]$ from those equations, we get two linear and coupled equations for $L[\delta\alpha]$ and $L[\delta\alpha^*]$ given by
	\begin{align}
		{\bf C}(s)\Vec{\textbf{b}}(s)=\Vec{\textbf{x}}(s)\label{coupledfield}
	\end{align}
	where we have introduced 
	\begin{equation}
		\begin{aligned}
			\Vec{\textbf{b}} &= \begin{pmatrix} L[\delta\alpha] \\ L[\delta\alpha^{*}] \end{pmatrix},\\
			\Vec{\textbf{x}} &= \begin{pmatrix} \delta\alpha(0) - iNg_cZ(s) \\ \delta\alpha^{*}(0) + iNg_cZ^{*}(s) \end{pmatrix},
		\end{aligned}
	\end{equation} 
	and the $2\times 2$ coupling matrix ${\bf C}$ with entries ${\bf C}_{ab}$ [$a,b\in\{1,2\}$] given by
	\begin{equation} \label{elementsofC}
		\begin{aligned}
			\mathcal{\textbf{C}}_{11}(s) &= s + i\Delta_c' + \frac{\kappa}{2} + Ng^{2}_cY(s) = \mathcal{\textbf{C}}_{22}^* (s),
		\end{aligned}
	\end{equation}	
	and
	\begin{equation}
	    \begin{aligned}
			\mathcal{\textbf{C}}_{12}(s) &= Ng^{2}_cX(s) = \mathcal{\textbf{C}}_{21}^* (s).
		\end{aligned}
	\end{equation}
	Here, we have defined
	\begin{equation} \label{XYZ}
		\begin{aligned}
			X(s) &= \Tr{\hat{\sigma}_{ge}W^{-1}(s)[\hat{\sigma}_{ge},\hat{\rho}_{ss}]}, \\
			Y(s) &= \Tr{\hat{\sigma}_{ge}W^{-1}(s)[\hat{\sigma}_{eg},\hat{\rho}_{ss}]}, \\
			Z(s) &= \Tr{\hat{\sigma}_{ge}W^{-1}(s)\delta\hat{\rho}(0)},
		\end{aligned}
	\end{equation}
	and 
	\begin{equation} \label{W}
		\begin{aligned}
			W(s) &= s - \hat{\mathcal{L}}_A - \hat{\mathcal{L}}_F[\alpha_{ss}].
		\end{aligned}
	\end{equation}
	Details of this derivation have been shifted to Appendix~\ref{LaplaceAppendix}. Equation~\eqref{coupledfield} can now be solved by inverting ${\bf C}(s)$ for every value of $s$.

\begin{figure}
		\centerline{\includegraphics[width=1\linewidth]{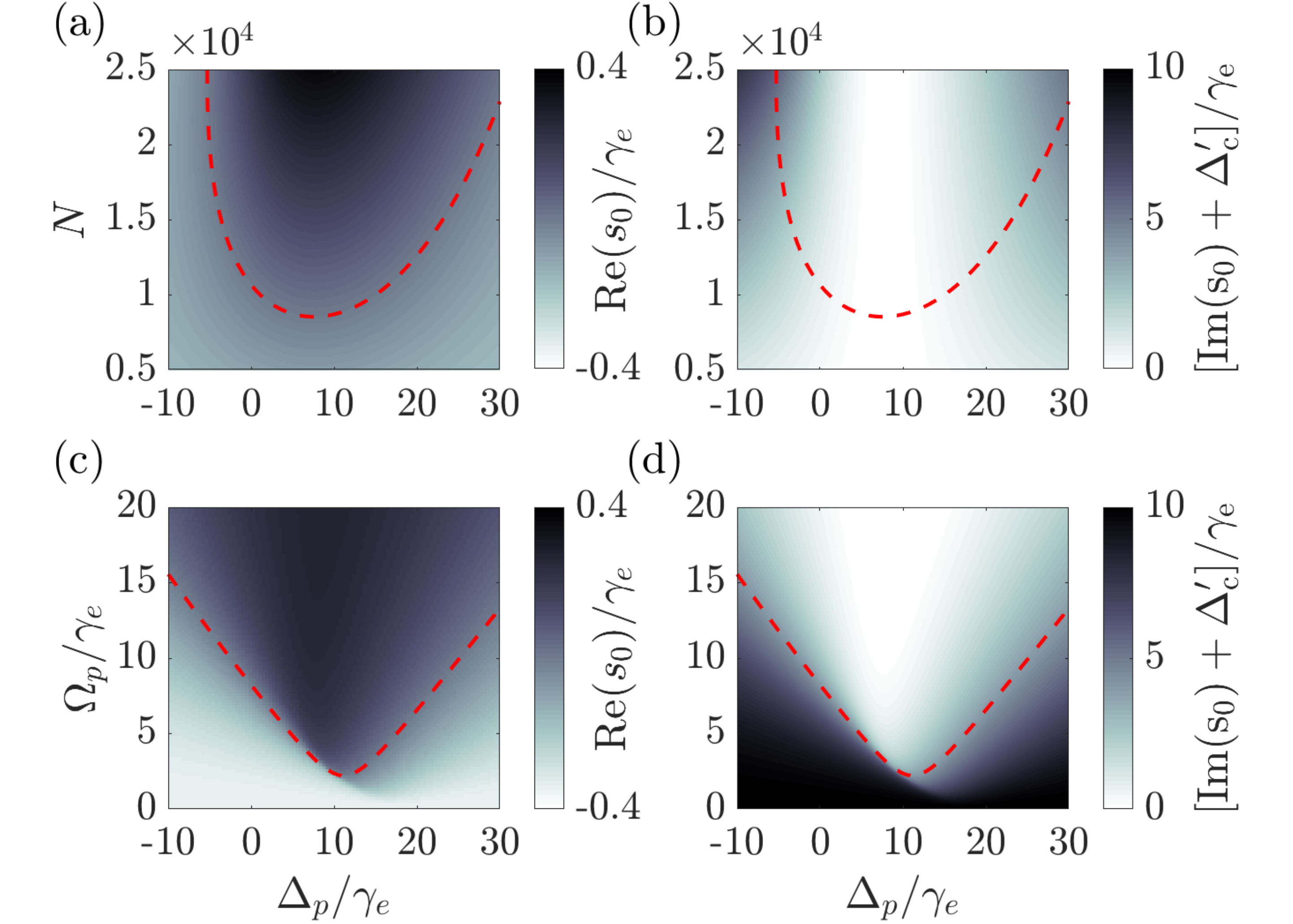}}
		\caption{(a) The real component $\Re ( s_0 )$ and (b) imaginary component $\Im ( s_0 )$ of the primary zero $s_0$, which possesses the largest real component of $D(s)$ in Eq.~\eqref{Dispersionrelation}, as a function of the pump detuning and atom number. 
			(c)--(d) The real and imaginary part of the zero $s_0$, respectively, as a function of the pump Rabi frequency $\Omega_p$ and detuning $\Delta_p$. 
			The red dashed line is the lasing threshold in which $\Re ( s_0 ) = 0$. 
			The common parameters for all plots are $\Delta_c = -192 \gamma_e$, $\Delta_m = -192 \gamma_e$, $g_c = 0.33 \gamma_e$, $\gamma_a = 159 \gamma_e$, $\Omega_m = \gamma_a/2$, and $\kappa = 0.39 \gamma_e$. Meanwhile, (a--b) has $\Omega_p = \sqrt{140} \gamma_e$ and (c--d) has $N = 20000$.}
		\label{Fig:2}
	\end{figure}

	The stability of the non-lasing solution $(\hat{\rho}_{ss},\alpha_{ss})$ is determined by whether $\delta \alpha$ is exponentially damping (stable) or exponentially growing (unstable). Stability for the fields thus require that all poles $s_n$ of the Laplace transformed fields $\Vec{\textbf{b}}(s)$ have a negative real part. 
	This is true since such a pole $s_n$ results in a field $\delta \alpha \propto e^{s_n t}$. To determine the stability, it is then sufficient to find the primary solution $s_0$ with the largest real component. Before finding $s_0$, we first mention that an instability cannot occur from a pole of $Z(s)$. This is because all values of $s$ for which $W(s)$ is not invertible are negative, which is equivalent to the statement that the spectrum of $\hat{\mathcal{L}}_A + \hat{\mathcal{L}}_F[\alpha_{ss}]$ consists of numbers with a negative real part. Then, the only way to find an instability is by a pole coming from inverting ${\bf C}(s)$. These poles can be found as the roots of the determinant of ${\bf C}(s)$, which is called the dispersion relation 
	\begin{equation} \label{Dispersionrelation}
		\begin{aligned}
			{D}(s)= \det \left[ \textbf{C} (s) \right].
		\end{aligned}
	\end{equation}
	Using this result, we can now numerically find $s_0$ by calculating the zero with the largest real part of Eq.~\eqref{Dispersionrelation}. We plot the real and imaginary parts of this primary root in Fig.~\ref{Fig:2}.
	The special case of $\Re \left( s_0 \right) = 0$ is the threshold value of the lasing transition, which we calculate numerically and display as a red dashed line in Fig.~\ref{Fig:2}.
	When $\Re \left( s_0 \right) < 0$, the non-lasing solution is stable and $\Re \left( s_0 \right)$ determines the decay rate of the fluctuations. Meanwhile, the imaginary part $\Im \left( s_0 \right)$  determines the frequency of the light emission as $\delta \alpha \propto \exp[{i \Im \left( s_0 \right) t}]$. 
	If $\Re \left( s_0 \right) > 0$, we expect an exponential increase in the field fluctuations, indicating that the non-lasing solution was unstable.
	
	It might be interesting to compare the mean-field results in Fig.~\ref{Fig:2} with our simplified threshold $\langle \hat{\sigma}_{ee}\rangle=\kappa/(N\gamma_{\mathrm{eff}})\approx3\times10^3/N$ that we introduced in Sec.~\ref{LasingMechanism}. For large pump power, we expect $\langle \hat{\sigma}_{ee}\rangle\approx0.5$ resulting in a threshold at $N\approx 6000$. This is in fact close to the lower bound of the threshold (red dashed line) visible in Figs.~\ref{Fig:2}(a) and (b). The curvature of the red line is likely due to $\langle \hat{\sigma}_{ee}\rangle\approx0.5$ being violated if the pump laser becomes off-resonant. It might seem surprising that the lowest critical value of $N$ is not found at $\Delta_p=0$. However, this can be partially explained by the existence of an AC-Stark shift that is induced by the off-resonant MOT laser. Since the MOT laser is red detuned from the $\ket{g}\leftrightarrow\ket{a}$ transition, it shifts the energy of the $\ket{g}$ state relative to the $\ket{e}$ state down by an amount 
		\begin{equation}
	\Delta_{{\rm AC},{\rm MOT}} \approx -\frac{\Omega_m^2}{4 \Delta_m},
	\end{equation}
	which is $\Delta_{{\rm AC},{\rm MOT}} \approx8.23\gamma_e$ for our parameters. We can then compensate for this shift by using a blue detuned pump laser.
	
	A similar shift is also found in Figs.~\ref{Fig:2}(c) and (d). Here, the threshold line seems to be nearly symmetric with respect to its minimum. Such a behavior is expected, according to our considerations in Sec.~\ref{LasingMechanism}. In fact, the critical pump power in this picture is principally determined by the requirement to pump enough atoms into the $\ket{e}$ state. The population of this state approaches $0.5$ for a diverging value of $\Omega_p^2/([\Delta_p-\Delta_p^{\mathrm{opt}}]^2+\Gamma^2/4)$. Here, $\Gamma$ is the effective linewidth of $\ket{e}$ and $\Delta_p^{\mathrm{opt}}$ accounts for all frequency shifts. Thus, we would expect the critical value of $\Omega_p$ to scale with $\sqrt{[\Delta_p-\Delta_p^{\mathrm{opt}}]^2+\Gamma^2}$, which is symmetric in $\Delta_p$ around $\Delta_p^{\mathrm{opt}}$ and also explains the linear slope of the red line in Figs.~\ref{Fig:2}(c) and (d) for large detuning $|\Delta_p|$. The asymmetry of the transition line in Figs.~\ref{Fig:2}(a) and (b) is likely due to a dependence of $\Delta_p^{\mathrm{opt}}$ and $\Gamma$ on $N$, which was also studied in Ref.~\cite{Gothe:2019}.

	Finally, we want to discuss the results of $\mathrm{Im}(s_0)$ that are visible in false colors in Figs.~\ref{Fig:2}(b) and (d). We find that close to the lasing threshold, the frequency of the amplified light field is $\mathrm{Im}(s_0)\approx-\Delta_c'$. Since all equations are reported in a reference frame where $\hat{c}$ rotates with $-\omega_p$, this means that the light is emitted approximately in resonance with $\omega_c$ in the lab frame. Nevertheless, we find a non-negligible shift of the light emitted into the cavity, which can be far detuned from the cavity resonance with respect to the cavity linewidth.

	\subsection{Floquet method} \label{FloquetMethod}
	While the stability analysis offers insight into the onset of the lasing dynamics, it inherently assumes an underlying non-lasing solution. The stability analysis can be used to calculate the lasing threshold. However, it cannot be used to calculate the actual lasing intensity and frequency at steady state.
	
	To find a description for the lasing solution, we now employ a Floquet method. We assume that the field and atomic density matrix can be decomposed in components corresponding to multiples of the frequency $\omega$. The frequency $\omega$ has to be found self-consistently.
	We first make a Fourier decomposition of the field and atomic density matrix into $2m + 1$ components for some cutoff frequency $\omega_{\text{cut}} = m \omega$. 
	The decomposition of $\alpha$ and $\hat{\rho}$ is a sum of time-independent amplitudes given by
	\begin{equation} \label{Fouierdecomp}
		\begin{aligned}
			\alpha = \sum_{n=-m}^{m} \alpha_{n}e^{i\omega nt},\\ \hat{\rho} = \sum_{n=-m}^{m} \hat{\rho}_{n}e^{i\omega nt}.  
		\end{aligned}
	\end{equation}\\
	Substituting this into Eqs.~\eqref{Fluctuationequationrho} and~\eqref{Fluctuationequationalpha} results in
\begin{equation}
		\begin{aligned}
			i\omega n\hat{\rho}_n &= \hat{\mathcal{L}}_A\hat{\rho}_n + \sum_{n^{\prime}=-m}^{m} \left( \hat{\mathcal{G}}_u[\alpha_{n - n^{\prime}}] +\hat{\mathcal{G}}_d[\alpha^*_{n^{\prime} - n}]\right)\hat{\rho}_{n^{\prime}},
        \end{aligned}
\end{equation}
and
\begin{equation}
    \begin{aligned}
			i\omega n\alpha_n &= -\left(i\Delta_c' + \frac{\kappa}{2}\right) \alpha_n - iNg_c\Tr{\hat{\sigma}_{ge}\hat{\rho}_n}, \label{Fixedfreqsolution}
	\end{aligned}
\end{equation}
	where we have introduced $\hat{\mathcal{G}}_u[\alpha]\hat{\rho}=-ig_c\alpha[\hat{\sigma}_{eg},\hat{\rho}]$ and $\hat{\mathcal{G}}_d[\alpha^*]\hat{\rho}=-ig_c\alpha^*[\hat{\sigma}_{ge},\hat{\rho}]$ and decomposed the coupling of the atomic density matrix with the field as $\hat{\mathcal{L}}_F[\alpha]=\hat{\mathcal{G}}_u[\alpha]+\hat{\mathcal{G}}_d[\alpha^*]$.
	
	The non-lasing solution $( \hat{\rho}_{ss}, \alpha_{ss})$, whose stability we have analyzed in Sec.~\ref{StabilityAnalysis}, can be understood as a limiting case of Eq.~\eqref{Fixedfreqsolution} where we impose $\hat{\rho}_m=0=\alpha_n$ for $m,n\neq0$. This is the case when there is, to good approximation, no additional field in the cavity except for the scattered laser light given by $\alpha_0$ [see Eq.~\eqref{alphasteadystate}]. Since this non-lasing solution becomes unstable, we expect to observe a component in the sideband $\alpha_{1}$ where the frequency $\omega$ is close to $-\Delta_c'$. This is by far the largest component and all other components are suppressed due to the small cavity linewidth that is much larger than the emission frequency $\kappa\ll\omega$. 
	
	Consequently, to a good approximation, we can consider only three frequency components, which is equivalent to performing a cut-off at $m = 1$.  Imposing this cutoff onto Eq.~\eqref{Fixedfreqsolution}, and therefore disregarding higher and lower frequency terms, allows us to rewrite Eq.~\eqref{Fixedfreqsolution} as $\boldsymbol{\mathfrak{L}}(\Vec{\boldsymbol{\alpha}},\omega) \Vec{\boldsymbol{\rho}} = \Vec{\textbf{0}}$, where $\Vec{\boldsymbol{\rho}} = (\hat{\rho}_{-1}, \hat{\rho}_0, \hat{\rho}_1)^T$, $\Vec{\boldsymbol{\alpha}}=({\alpha}_{-1}, {\alpha}_0, {\alpha}_1)^T$, and
	\begin{equation}
		\begin{aligned}
			\boldsymbol{\mathfrak{L}}(\Vec{\boldsymbol{\alpha}},\omega) = \begin{pmatrix}
				-i\omega - \mathfrak{\hat{L}}_0 & -\mathfrak{\hat{L}}_{-1} & 0\\ \\
				-\mathfrak{\hat{L}}_1 &-\mathfrak{\hat{L}}_0  & -\mathfrak{\hat{L}}_{-1} \\ \\
				0 & -\mathfrak{\hat{L}}_1 & i\omega - \mathfrak{\hat{L}}_0
			\end{pmatrix}.
		\end{aligned}
	\end{equation}
	The elements of $\boldsymbol{\mathfrak{L}}(\Vec{\boldsymbol{\alpha}},\omega)$ are given by
	\begin{equation}
		\begin{aligned}
			\mathfrak{\hat{L}}_{-1} &= \hat{\mathcal{G}}_u[\alpha_{-1}] + \hat{\mathcal{G}}_d[\alpha^*_1],
	\end{aligned}
	\end{equation}
	\begin{equation}
		\begin{aligned}
			\mathfrak{\hat{L}}_0 &= \hat{\mathcal{L}}_A + \hat{\mathcal{G}}_u[\alpha_0] + \hat{\mathcal{G}}_d[\alpha^*_0],
	    \end{aligned}
	\end{equation}
	and
	\begin{equation}
		\begin{aligned}
			\mathfrak{\hat{L}}_1 &= \hat{\mathcal{G}}_u[\alpha_{1}] + \hat{\mathcal{G}}_d[\alpha^*_{-1}].
		\end{aligned}
	\end{equation}
	We can now find the steady state $\Vec{\boldsymbol{\rho}}(\Vec{\boldsymbol{\alpha}},\omega)$ by calculating the kernel of $\boldsymbol{\mathfrak{L}}(\Vec{\boldsymbol{\alpha}},\omega)$ and imposing the normalization condition. This steady state depends on the choice of the field $\Vec{\boldsymbol{\alpha}}$ and $\omega$ which has to be updated self-consistently. To do this, we calculate $\Vec{\tilde{\boldsymbol{\alpha}}}(\Vec{\boldsymbol{\alpha}},\omega)=(\tilde{\alpha}_{-1}(\Vec{\boldsymbol{\alpha}},\omega), \tilde{\alpha}_0(\Vec{\boldsymbol{\alpha}},\omega), \tilde{\alpha}_1(\Vec{\boldsymbol{\alpha}},\omega))^T$ with
	\begin{align}
	\tilde{\alpha}_n(\Vec{\boldsymbol{\alpha}},\omega)=\frac{-iNg_c\mathrm{Tr}\{\hat{\sigma}_{ge}\hat{\rho}_n(\Vec{\boldsymbol{\alpha}},\omega)\}}{i(\Delta_c'+\omega n)+\frac{\kappa}{2}},
	\end{align}
	where $\hat{\rho}_n(\Vec{\boldsymbol{\alpha}},\omega)$ is the $n$ component of the steady-state vector $\Vec{\boldsymbol{\rho}}(\Vec{\boldsymbol{\alpha}},\omega)$. Then, we define a function
	\begin{align}
	F(\Vec{\boldsymbol{\alpha}},\omega)=\sum_{n=-m}^m|\alpha_n-\tilde{\alpha}_{n}(\Vec{\boldsymbol{\alpha}},\omega)|^2,\label{F}
	\end{align}
	such that $F(\Vec{\boldsymbol{\alpha}},\omega)=0$ results in the realization of a steady state for the field $\Vec{\boldsymbol{\alpha}}$, the state of the atom $\Vec{\boldsymbol{\rho}}$ and the frequency $\omega$. 
	\begin{figure} 
		\centerline{\includegraphics[width=\linewidth]{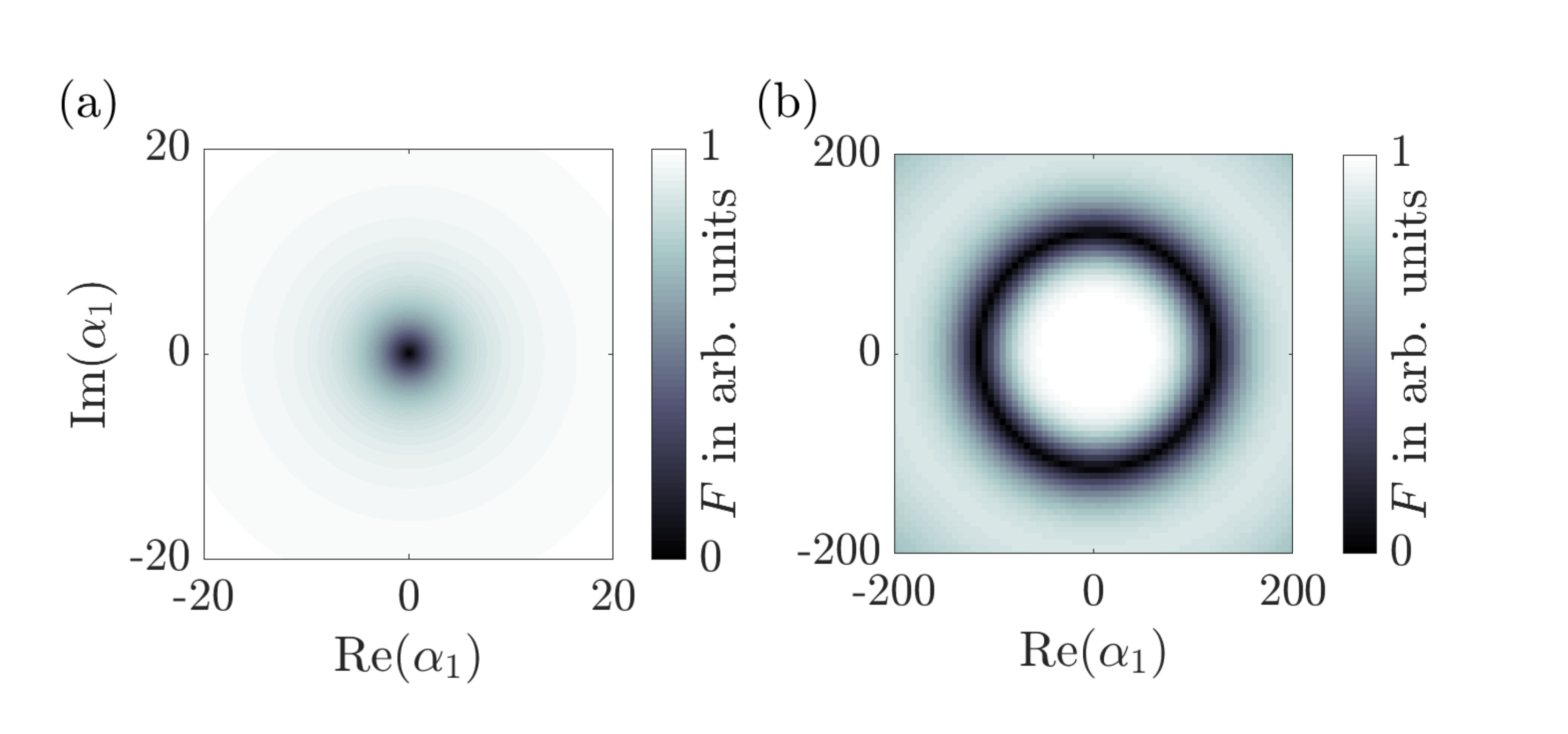}}
		\caption{(a) The function $F$ defined in Eq.~\eqref{F} in arbitrary units  for $N = 20000$, $\Delta_p = 10\gamma_e$, $\Omega_p = 2\gamma_e$, , $\alpha_{-1}=0$ $\alpha_0 \approx -3.3 - 2.6i$, and $\omega >0$. (b) The function $F$ for $\Omega_p = 15\gamma_e$, $\alpha_{-1} \approx 0$, $\alpha_0 \approx -5.6 - 1.3i$, and $\omega \approx -205.5\gamma_e$.}
		\label{Fig:3}
	\end{figure}
	
	Finding a solution $\Vec{\boldsymbol{\alpha}},\omega$ of Eq.~\eqref{F} is achieved numerically and we show $F(\alpha_{-1},\alpha_0,\alpha_1,\omega)$ in the non-lasing (a) and lasing regime (b) in Fig.~\ref{Fig:3} for a fixed choice of $\alpha_0,\alpha_{-1}$ and $\omega$ [see inset of Fig.~\ref{Fig:3}]. In the non-lasing case, we find that the only zero is found at $\alpha_1=0$ as seen in Fig.~\ref{Fig:3}(a). This indicates that the only light field in the cavity is the scattered laser light by the atoms given by $\alpha_0$. In the lasing regime shown in Fig.~\ref{Fig:3}(b), we find a $U(1)$ symmetric set of solutions indicating a non-vanishing lasing field amplitude $|\alpha_1|^2$ with an arbitrary phase. This $U(1)$ symmetry is a direct consequence of the underlying equations that are invariant under a transformation $\alpha_1\mapsto\alpha_1\exp(-i\varphi)$, $\hat{\rho}_1\mapsto\exp(-i\varphi\hat{\sigma}_{ee})\hat{\rho}_1\exp(i\varphi\hat{\sigma}_{ee})$ with arbitrary phase $\varphi$. This transformation is defined in the Fourier components. Notice that the total field $\alpha$ and atomic density matrix $\hat{\rho}$ is not invariant under this transformation. The appearance of a $U(1)$ symmetric solution is a common feature of laser systems and highlights that the phase of the laser is spontaneously broken. In fact, this directly implies that the field $\alpha_1$ is not locked to the phase of an external driving laser.
	
	Using Eq.~\eqref{F} we find the solution $\Vec{\boldsymbol{\alpha}},\omega$ and calculate the total, time-averaged field
	\begin{equation}
	    \begin{aligned}
	\abs{\alpha}_{\text{av}}^2 &= \lim_{t_{\text{av}} \to \infty} \int_0^{t_{\text{av}}} \frac{dt}{t_{\text{av}}} \abs{\alpha}^2 \\
	&= \abs{\alpha_{-1}}^2 + \abs{\alpha_0}^2 + \abs{\alpha_1}^2.
	    \end{aligned}
	\end{equation}
	We are now in a position to reexamine the lasing threshold by studying the intensity of the field for different parameters using our three-component Floquet method.
	\begin{figure}
		\centerline{\includegraphics[width=1\linewidth]{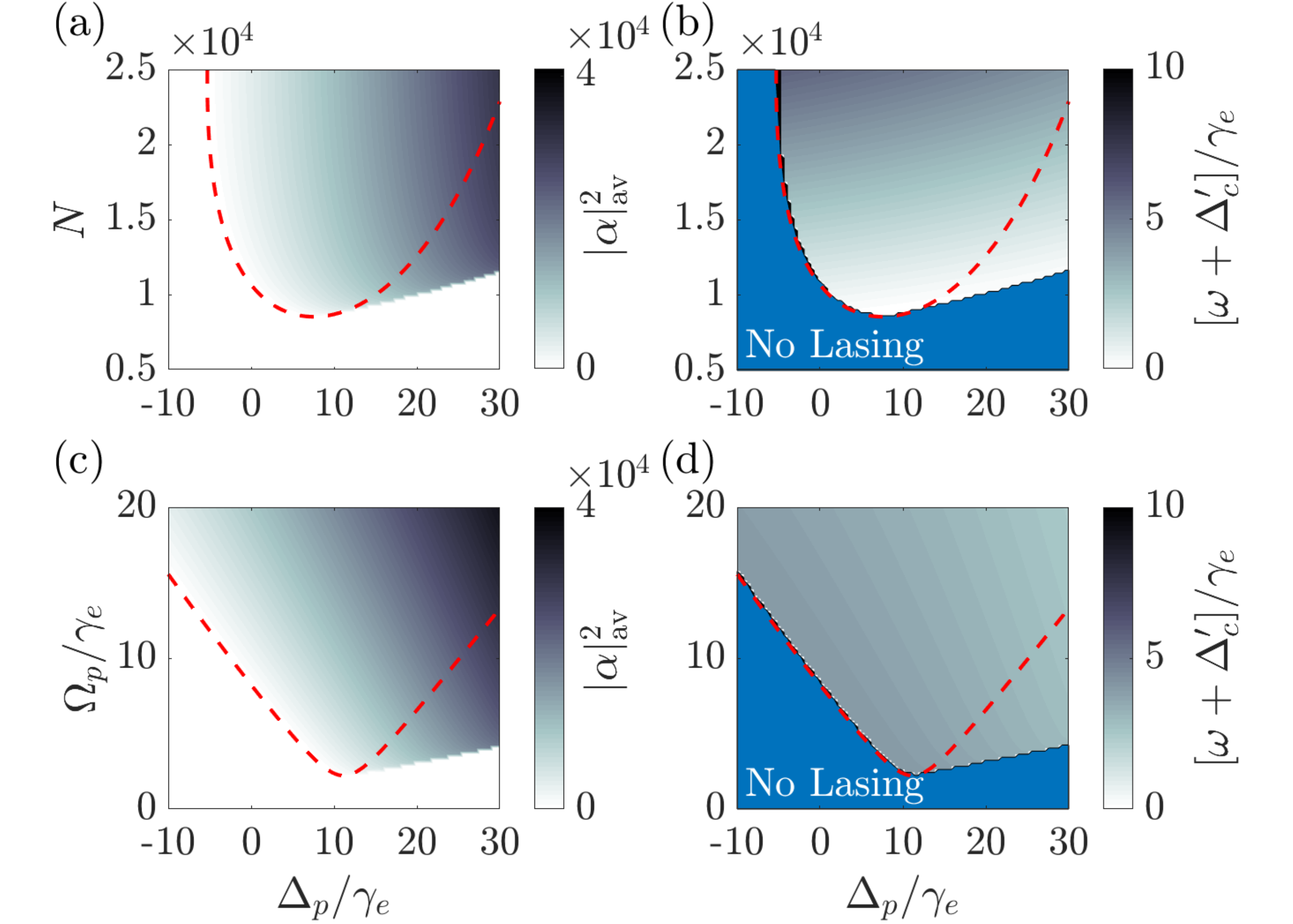}}
		\caption{Floquet method calculations of the intensity [(a) and (c)] and emission frequency [(b) and (d)] of the lasing field.
			The first row (a)--(b) show these values as a function of atom number and pump detuning while the second row plots (c)--(d) instead vary the pump Rabi frequency and pump detuning. 
			The red dashed lines were calculated using Eq.~\eqref{Dispersionrelation}. 
			All parameters are the same as in Fig.~\ref{Fig:2}.}
		\label{Fig:4}
	\end{figure}
	We do this in Fig.~\ref{Fig:4} where we examine the same parameter regimes as Fig.~\ref{Fig:2} and display the threshold found using the stability analysis as a red dashed line in each plot. The intensities are visible in Fig.~\ref{Fig:4}(a) and (c). For negative and small values of $\Delta_p$, we find that the red dashed line is in good agreement with the onset of a large cavity field. Instead, interestingly, we find that there exists a region for large $\Delta_p$ in which the stability analysis suggests that the non-lasing solution is stable and yet our Floquet method predicts a large intracavity photon number. This suggests that there exists a bistability and a coexistence of a lasing and a non-lasing solution.

	In Fig.~\ref{Fig:4}(b) and (d), we show the emission frequency~$\omega$ for the same parameters. In the case where we found $\alpha_{1} = 0$, this frequency is not defined, which we indicate in these plots as the `No Lasing' regime. In the lasing regime, we find the frequency $\omega$ to be close, but slightly detuned, from the cavity resonance $-\Delta_c'$. This shows that there are non-trivial light-shifts that modify the lasing frequency. As visible in Fig.~\ref{Fig:4}(d), this frequency seems to be almost independent of $\Omega_p$ and $\Delta_p$, while we see a major dependence of this frequency $\omega$ on the atom number $N$ [see Fig.~\ref{Fig:4}(d)]. Here, we see an increase of the detuning $\omega+\Delta_c'$ with the number of atoms~$N$.

	\subsection{Hysteresis regime} \label{HysteresisRegime}
\begin{figure}
		\centerline{\includegraphics[width=\linewidth]{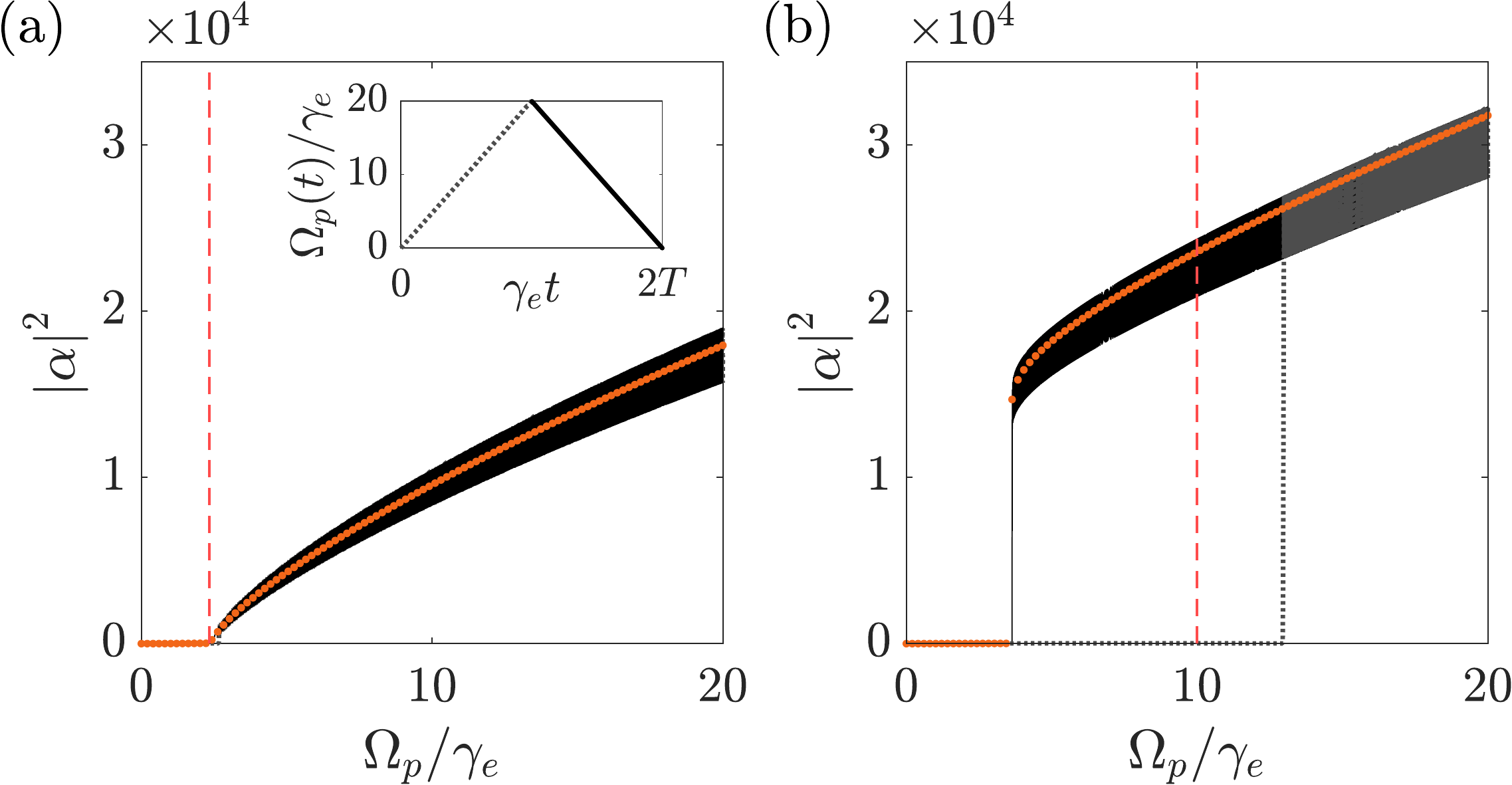}}
		\caption{Intensity $\abs{\alpha}^2$ as a function of $\Omega_p = A t$ for $20000$ atoms. 
			The evolution uses (a) $\Delta_p = 10\gamma_e$ which should not possess a hysteresis regime and (b) $\Delta_p = 25\gamma_e$ which has a hysteresis regime for $4 \gamma_e \lesssim \Omega_p \lesssim 10 \gamma_e$. 
			In both plots, the gray curve represents the `forward' evolution as $\Omega_p$ increases linearly, while the black curve represents the `backwards' evolution with $\Omega_p$ decreasing linearly, with ramping rate $A \approx 3.1 \times 10^{-4}\gamma_e^2$ and integration time $T = 64000/\gamma_e$. 
			Meanwhile, the red dashed curves are the lasing thresholds calculated from the stability analysis, i.e., when $Re( s_0 ) = 0$. 
			The orange dotted curves display the value of $\abs{\alpha}^2$ calculated from the MFFM for a particular pump power.
			The inset displays the forward and backwards ramping of $\Omega_p$.}
		\label{Fig:5}
	\end{figure}

	Now, we want to study the regime where we expect to have bistable non-lasing and lasing solutions. We do this dynamically by ramping the pump power up and down again. 
	We initialize the system with $\expval{\hat{\sigma}_{gg}^{(j)} (t = 0)} = 1$ and $\abs{\alpha (t = 0)}^2 \approx 0$, and require that the pump laser is initially off $\Omega_p (t = 0) = 0$. This is in the non-lasing regime when the cavity field is basically empty. Using this initial state, we simulate the dynamics of the field and the mean-field atomic density operator whose evolution is governed by Eqs.~\eqref{alphaequationofmotion} and~\eqref{Meanfieldmasterequation}.
	While integrating those equations, we sweep the pump Rabi frequency with a linear profile $\Omega_p (t) = A t$ with a slow rate $A$ until $\Omega_p = 20 \gamma_e$ at time $T$. This value is chosen such that, for all parameters visible in Fig.~\ref{Fig:4}, we end up with only the lasing solution. After this ramping up of the power, we ramp down the power with the same but negative linear slope $-A$ such that $\Omega_p(t)=20\gamma_e-A(t-T)$ and $\Omega_p(2T)=0$. We show the ramping scheme as a sketch, in the inset of Fig.~\ref{Fig:5}.
    
	We perform these simulations for two different detunings representing the parameters where we do not expect to find bistability $\Delta_p = 10 \gamma_e$ [Fig.~\ref{Fig:5}(a)] and where we expect to find bistability $\Delta_p = 25 \gamma_e$ [Fig.~\ref{Fig:5}(b)]. In Fig.~\ref{Fig:5}(a) and (b) we plot the dynamics of the field intensity $|\alpha(t)|^2$ for ramping up the power as gray dotted lines and for ramping down the power as a black solid line. As visible in Fig.~\ref{Fig:5}(a), the black line completely overlaps with the gray data point therefore indicating that the lasing solution is the same when ramping up or down the pump power. We also compare the dynamically simulated laser intensity with $|\alpha|_{\mathrm{av}}$ obtained from the MFFM visible as orange dots. We find excellent agreement showing that we adiabatically track the lasing solution. 
	
	In Fig.~\ref{Fig:5}(b), we see a quite different behavior due to the existence of bistability. When ramping up the laser power, the atomic system starts lasing for a pump power that is even beyond the one predicted by the stability analysis (vertical dashed line). We expect that this is due to the fact that although our ramp speed $A$ is slow, it can never be adiabatic when crossing a transition. When ramping the power down, we find that in the regime where we previously found only a very small cavity field, we are now in a lasing regime. The light intensity then suddenly jumps to zero for sufficiently weak pumping $\Omega_p\lesssim4\gamma_e$. This is in agreement with the MFFM (orange dots). Our finding is a clear indication of hysteresis in this atom-cavity system which is highlighting the non-linear nature of the atom-light coupling.
	
    \subsection{Discussion of the mean-field results}
		\label{Comparison:MF-SOC}
	Our analysis treats both atom-atom and atom-cavity interactions at a mean-field level. In addition, it is assumed that the field is in a `classical' coherent state which is often a good approximation in laser theory. However, the fact that we have disregarded the effects of fluctuations and correlations only allows us to derive certain properties such as the laser intensity and frequency of the cavity field. In the following, we want to benchmark our results with a second-order cumulant approximation which includes fluctuations in the atomic and cavity degrees of freedom to a certain extent.
	
	The second-order cumulant description is derived by calculating the time derivative of all mean-field values~$\langle \hat{A}\rangle$ and second-order moments $\langle\hat{A}\hat{B}\rangle$ where $\hat{A},\hat{B}$ are arbitrary single atom operators or the cavity field operators $\hat{c}, \hat{c}^{\dagger}$. Taking advantage of the permutation symmetry and factorizing third-order moments, one can then find a closed set of equations for all first $\langle \hat{A}\rangle$ and second-order moments $\langle\hat{A}\hat{B}\rangle$, where all single atom operators are for atoms $j = 1,2$. The exact derivation is described in Ref.~\cite{Plankensteiner:2022} and also applied to a similar system in Ref.~\cite{Hotter}. Since the cumulants also include the dynamics of second-order moments, they are considered to go beyond the mean-field description that is presented in this paper.
	
\begin{figure}
		\centerline{\includegraphics[width=\linewidth]{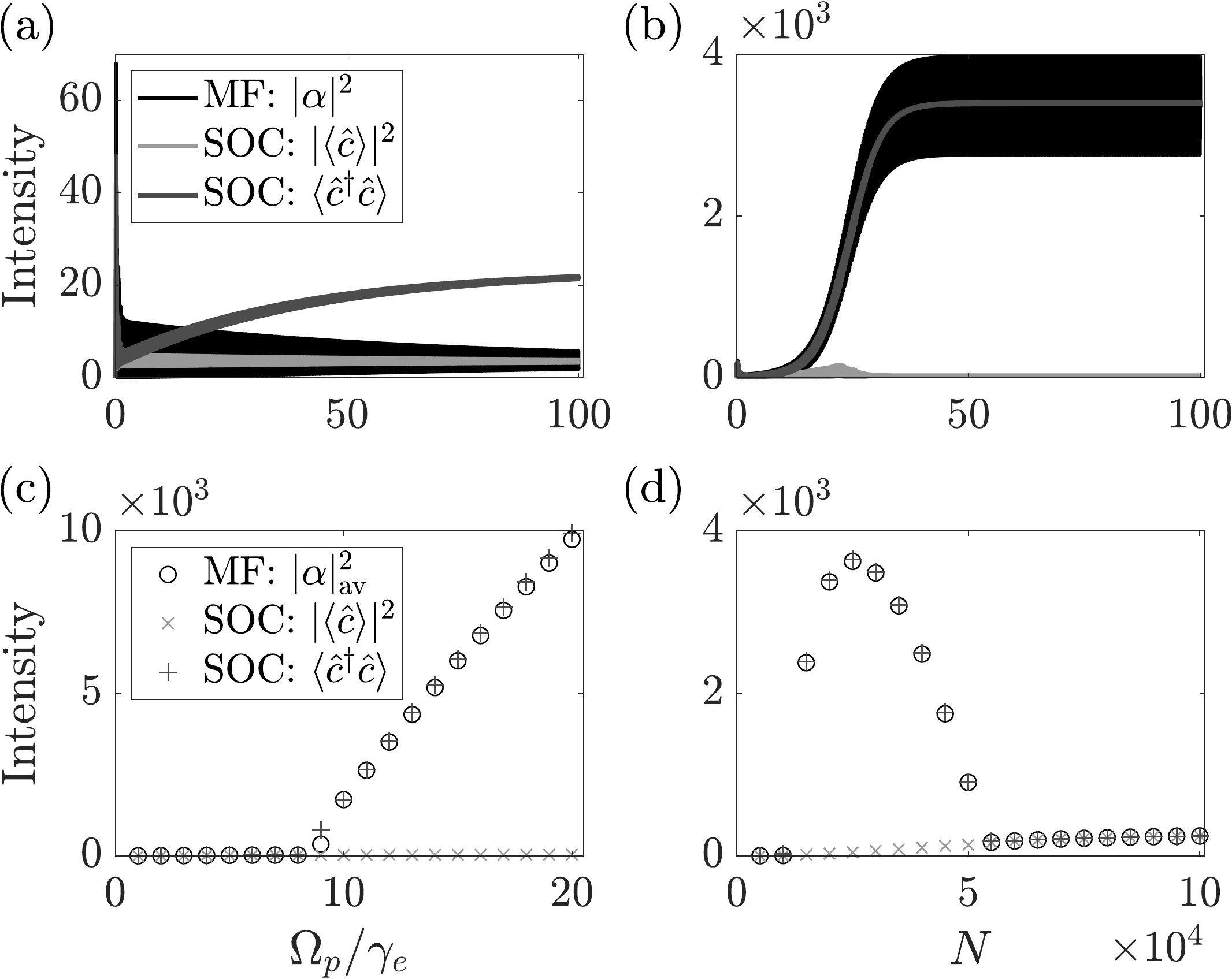}}
		\caption{Dynamical comparison of the intensity $\abs{\alpha}^2$ for mean-field (MF, blue curves) and $\expval{\hat{c}^{\dagger} \hat{c}}$ for second-order cumulants (SOC, red curves).
			Also shown is the coherent component of the field $\abs{\expval{c}}^2$ from the second-order code (orange curves).
			The evolution uses the same parameters as Fig.~\ref{Fig:2} expect: (a) $N = 10000$, $\Omega_p = \sqrt{140} \gamma_e$, $\Delta_p = 0$; (b) $N = 20000$, $\Omega_p = \sqrt{140} \gamma_e$, $\Delta_p = 0$; Steady-state intensities (blue circles and red pluses) and coherent field (orange crosses) as a function of $\Omega_p$.
			The parameters are the same as in Fig.~\ref{Fig:2} with $N = 20000$ and $\Delta_p = 0$.}
		\label{Fig:6}
	\end{figure}

	We now want to compare the second-order cumulant approximation with our mean-field results. In a first comparison, we analyze the dynamics for exemplary parameters above and below the lasing threshold. In Fig.~\ref{Fig:6}(a), we show the dynamics of $|\alpha|^2$ calculated from the mean-field results visible as black solid line. The dynamics is shown for $\Omega_p=\sqrt{140}\gamma_e$, $\Delta_p=0$, and $N=10000$, which is slightly below the lasing transition visible in Fig.~\ref{Fig:2}. We compare the dynamics of this mean-field trajectory with the ones of the second-order cumulants results for $|\langle\hat{c}\rangle|^2$ and $\langle\hat{c}^{\dag}\hat{c}\rangle$. The mean-field predicts a very similar trajectory as the one of $|\langle\hat{c}\rangle|^2$ while  $\langle\hat{c}^{\dag}\hat{c}\rangle$ is significantly higher. The reason for this is that the mean-field result $\alpha$ is mostly dominated by the coherently scattered laser field [see Eq.~\eqref{alphasteadystate}], which is also described as a coherent field in the second-order cumulants $\langle\hat{c}\rangle$. However, the second-order cumulants also describe the incoherent field $\langle\hat{c}^{\dag}\hat{c}\rangle-|\langle\hat{c}\rangle|^2$ which provides a major contribution to the total field below the lasing threshold.
	
	Above the lasing threshold, for $N=20000$ and the same pump power $\Omega_p=\sqrt{140}\gamma_e$ and detuning $\Delta_p=0$, we show the dynamics of the mean-field and second-order cumulant description in Fig.~\ref{Fig:6}(b). We now observe that the mean-field trajectory $\alpha$ oscillates around the mean intensity $\langle\hat{c}^{\dag}\hat{c}\rangle$ of the second-order cumulant description. The coherent field amplitude $|\langle\hat{c}\rangle|^2$ of the second-order cumulant description is instead very small. Here, the scattered laser field is just a minor part of the total light field which explains the small value of $|\langle\hat{c}\rangle|^2$. Instead, the lasing field that oscillates approximately at the cavity resonance is much more intense. This lasing field is described completely differently by the mean-field and the second-order cumulant descriptions. For the mean-field theory, this lasing field is purely coherent and achieved by breaking an underlying $U(1)$ symmetry. Our mean-field approach assumes a vanishing linewidth of this lasing field which is an artifact of our approach whose origin is the disregard of noise. The second-order cumulant description includes noise to a certain extent, and therefore the lasing field has a finite linewidth. The latter results in the fact that this lasing field is described as a incoherent component. Notice that the lasing component in mean-field $|\alpha_1|^2\approx\langle\hat{c}^{\dag}\hat{c}\rangle-|\langle\hat{c}\rangle|^2$ is approximately the incoherent light field in the second-order cumulant theory. This analysis shows the power and also the limitations of the mean-field theory described here. The mean-field theory can describe the lasing intensity and dynamics, however, it completely fails to describe the coherence time of this lasing field and instead assumes it is coherent on arbitrary timescales.

	To compare the second-order cumulant results of the intensity with the one of the mean-field description, we show different thresholds as a function of $\Omega_p$ and $N$ in Fig.~\ref{Fig:6}(c) and (d). We find good agreement of the time-averaged mean-field intensity $|\alpha|_{\mathrm{av}}^2$ and the second-order cumulant results $\expval{\hat{c}^{\dag}\hat{c}}$ across the non-lasing to lasing transition. Moreover, we observe that both, mean-field and second-order cumulants, predict a lasing transition at sufficiently large atom number $N$ and again a transition to non-lasing for even larger atom number $N$ [see Fig.~\ref{Fig:6}]. The second transition is explained by a increasing frequency shift for increasing atom number $N$ as it was also visible in the emission frequency of Fig.~\ref{Fig:4}(b). We expect that this shift eventually becomes too large to achieve enough population in $\ket{e}$ state and therefore leads again to a non-lasing configuration. The fact that this is described by both mean-field and second-order cumulant descriptions is a strong indicator for mean-field being reliable to describe the correct frequency shifts.

	\section{Conclusion} \label{Conclusion}

	In this paper, we have developed a theoretical model to study a lasing scheme for three level atoms in $V$-configuration coupled to an optical cavity.
	Our model employed a mean-field approximation of the cavity field and for the atomic operators which allowed us to simulate the large atom and intracavity photon numbers required to study the lasing transition.
	We preformed a stability analysis of a non-lasing solution, which allowed us to find the threshold for lasing and the initial emission frequency. We have analyzed these quantities in terms of changing the pump laser power and frequency as well as the total number of atoms. In addition, we were able to predict the intensity and frequency of the lasing solution using a Floquet analysis of the mean-field master equation and cavity field. This Floquet solution shows a $U(1)$ symmetry in one of the frequency components while the total atomic density matrix and cavity field do not possess a $U(1)$ symmetry. Furthermore, within this analysis we were also able to predict a bistable region that we tested by observing a hysteresis within our mean-field approach. Finally, we benchmarked our mean-field descriptions with results from a second-order cumulant theory and discussed its validity.
	
	We expect that the methods presented here can be extended in several ways. One possibility is to add noise in the cavity and atomic variables such that we can predict a finite, non-vanishing linewidth of the laser field. In addition, it might be interesting to study motion in this model. Motion can result in additional inhomogeneous broadening due to the Doppler shift of emitted photons which might well modify the lasing threshold and emission frequency. On the other hand, it might be also possible to control the motion of the atom in the lasing regime leading to cooling and trapping in coexistence with lasing~\cite{Xu:2016,Jaeger:2017,Hotter}. The realization of the latter would be intriguing as an example of a self-sustainable quantum device which produces coherent light and cools and traps the atoms at the same time.

	\section*{Acknowledgments}
	We would like to thank J. Cooper, A. Shankar, J. Bartolotta, J. Eschner, D. Sholokhov, S. Shaju, and C. Hotter for stimulating discussions.
	This research was supported by the NSF PFC Grant No. 1734006; NSF AMO Grant No. 1806827; and the NSF Q-SEnSE Grant No. OMA 2016244.
	
	\bibliography{references.bib}
	\appendix
	
	\section{Laplace Transformation of Field Fluctuations} \label{LaplaceAppendix}

	In this section we show additional steps which we used to calculate the dispersion relation given in Eq.~\eqref{Dispersionrelation}. Using Eqs.~\eqref{Fluctuationequationrho} and~\eqref{Fluctuationequationalpha}, the dynamics of $\delta\alpha(t)$ is governed by
	\begin{equation} \label{Fluctuationequationalphaappendix}
		\begin{aligned}
			\partial_t{\delta\alpha} &= -\left(i\Delta_c' + \frac{\kappa}{2} \right)\delta\alpha - iNg_c\Tr{\hat{\sigma}_{ge}\delta\hat{\rho}}.
		\end{aligned}
	\end{equation}
	The Laplace transformation Eq.~\eqref{laplace} of Eq.~\eqref{Fluctuationequationalphaappendix} leads to
	\begin{equation} \label{Laplacetransformalphaappendix}
		\begin{aligned}
			sL[\delta\alpha] &= \delta\alpha(0) - (i\Delta_c' + \frac{\kappa}{2})L[\delta\alpha]\\ &-iNg_c\Tr{\hat{\sigma}_{ge}L[\delta\hat{\rho}]}.
		\end{aligned}
	\end{equation}
	We look to solve for $L[\delta\alpha(t)]$ and thus need the Laplace transform of Eq.~\eqref{Fluctuationequationrho} which is
	\begin{equation} \label{Laplacetransformrhoappendix}
		\begin{aligned}
			L[\delta\hat{\rho}] &= W^{-1}(s)\delta\hat{\rho}(0) - ig_cL[\delta\alpha^{*}]W^{-1}(s)[\hat{\sigma}_{ge},\hat{\rho}_{ss}]\\ &-ig_cL[\delta\alpha]W^{-1}(s)[\hat{\sigma}_{eg},\hat{\rho}_{ss}],
		\end{aligned}
	\end{equation}
	where $W(s)^{-1}$ is the inverse of the operator given in Eq.~\eqref{W}. Then, after substituting Eq.~\eqref{Laplacetransformrhoappendix} into Eq.~\eqref{Laplacetransformalphaappendix}, we arrive at 
	\begin{equation}
		\begin{aligned}
			L[\delta\alpha] &= s^{-1}\Big[\delta\alpha(0) - \left(i\Delta_c' + \frac{\kappa}{2}\right)L[\delta\alpha]\\ &- iNg_cZ(s)- Ng^{2}_{c}L[\delta\alpha^{*}]X(s)\\ &- Ng^{2}_{c}L[\delta\alpha]Y(s) \Big].
		\end{aligned}
	\end{equation}
	Combining this equation with its conjugate results in the matrix relation Eq.~\eqref{coupledfield}. From Eq.~\eqref{coupledfield}, we derive can then the dispersion relation given by Eq.~\eqref{Dispersionrelation}.
\end{document}